\documentstyle[aps,prl,preprint,floats,epsfig]{revtex}

\begin{document}

\preprint{\tighten\vbox{\hbox{\hfil CLNS 01/1735}
                        \hbox{\hfil CLEO 01-9}
}}

\title{
First Observation of $\overline{B}^0\to D^{*0}\pi^+\pi^+\pi^-\pi^-$ Decays}
\author{CLEO Collaboration}
% You will want to hard code the date once you are ready to submit your paper!
\date{May 23, 2001}

\maketitle
\tighten

\begin{abstract} 
We report on the observation of  $\overline{B}^0\to D^{*0}\pi^+\pi^+\pi^-
\pi^-$ decays. The branching ratio is (0.30$\pm$0.07$\pm$0.06)\%. 
Interest in this particular mode
was sparked by Ligeti, Luke and Wise who propose it as a way to 
check the validity of factorization tests in 
$\overline{B}^0\to D^{*+}\pi^+\pi^-\pi^-\pi^0$ decays.  

\end{abstract}

\newpage
{\renewcommand{\thefootnote}{\fnsymbol{footnote}}
\begin{center}
K.~W.~Edwards,$^{1}$
A.~J.~Sadoff,$^{2}$
R.~Ammar,$^{3}$ A.~Bean,$^{3}$ D.~Besson,$^{3}$ X.~Zhao,$^{3}$
S.~Anderson,$^{4}$ V.~V.~Frolov,$^{4}$ Y.~Kubota,$^{4}$
S.~J.~Lee,$^{4}$
% ...
\end{center}
 
\small
\begin{center}
$^{1}${Carleton University, Ottawa, Ontario, Canada K1S 5B6 \\
and the Institute of Particle Physics, Canada}\\
$^{2}${Ithaca College, Ithaca, New York 14850}\\
$^{3}${University of Kansas, Lawrence, Kansas 66045}\\
$^{4}${University of Minnesota, Minneapolis, Minnesota 55455}\\
% ...
\end{center}

\setcounter{footnote}{0}
}
\newpage
%\section{Introduction}\label{sec:Introduction}
Factorization is the assumption that in two-body hadronic $B$ decays the decay 
amplitude can be expressed as a product of two currents, just as in
semileptonic decays where one current is hadronic and the other leptonic. 
Use of factorization has been crucial in creating models for understanding
the underlying weak decay dynamics \cite{factorization}.  

\begin{figure}[bht]
%\vspace{-2.5cm}
%\centerline{\epsfig{figure=Bto4pi.eps,height=4in}}
\centerline{\epsfig{figure=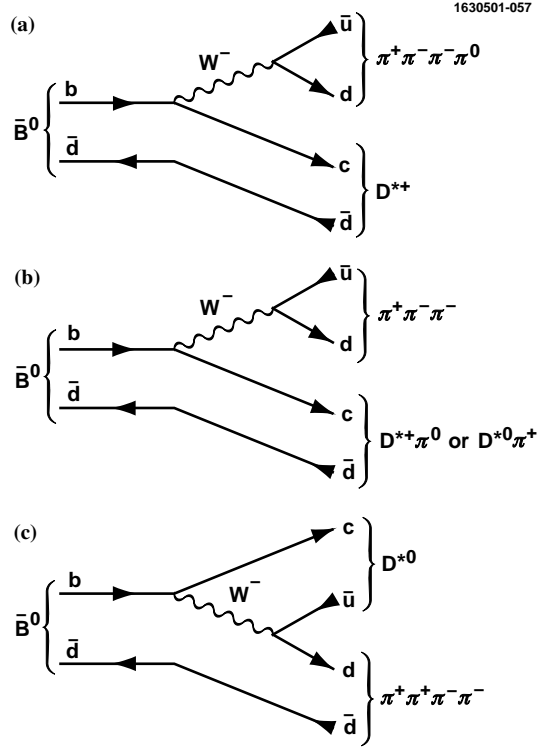,height=4in}}
%\vspace{-1.0cm}
\caption{\label{Bto4pi} Diagrams for $\overline{B}\to
D^*\pi\pi\pi\pi$ decays. (a) Charged current tree level diagram
for $D^{*+}(4\pi)^-$. (b) Charged current tree level
diagram for $\left(D^{*+}\pi^0\right)\pi^+\pi^-\pi^-$ or
$\left(D^{*0}\pi^+\right)\pi^+\pi^-\pi^-$. The $D^*\pi$ system
can form a $D^{**}$ resonance. (c) Color suppressed
diagram for $D^{*0}(4\pi)^0$.}
\end{figure}

In previous work we found a large branching fraction of (1.72$\pm$0.14
$\pm$0.24)\% for the
decay $\overline{B}^0\to D^{*+}\pi^+\pi^-\pi^-\pi^0$ \cite{Wangsbest}. 
This reaction can proceed via several possible tree level processes. 
The simplest diagram, shown in Fig.~\ref{Bto4pi}(a), has the four-pions emitted from
the virtual $W^-$. Assuming that this is indeed the dominant process, Ligeti,
Luke and Wise (LLW) \cite{LLW} have compared the $4\pi^-$ invariant mass spectrum  
with $\tau^-\to
\pi^+\pi^-\pi^-\pi^0\nu$ data \cite{CLEOtau}. Using a
model based on factorization they show that
the data agree up to a 4$\pi^-$ 
mass-squared of 2.9 GeV$^2$, within a precision of about 15\%. 

However, the agreement may be fortuitous, rather than a success of
factorization, if other diagrams are present. For example another possible
diagram is shown in Fig.~\ref{Bto4pi}(b), where the $D^{*+}$ and the $\pi^0$ are produced
at the lower vertex and the virtual $W^-$ manifests as $\pi^+\pi^-\pi^-$. This
process was searched for in the original publication. Definite evidence was
lacking but a stringent upper limit could not be set.

Here we search for the process $\overline{B}^0\to D^{*0}\pi^+\pi^+\pi^-\pi^-$
as suggested by LLW.
This can be produced by the diagram in Fig.~\ref{Bto4pi}(b), where the $D^{*0}$ combines
with one of the $\pi^+$'s to form a low-mass system. It could also be produced
by the color-suppressed process shown in Fig.~\ref{Bto4pi}(c).
In this paper we indeed show that the process
$\overline{B}^0\to D^{*0}\pi^+\pi^+\pi^-\pi^-$ has a significant branching
ratio and try to ascertain the dominant production mechanism.
The data sample consists of 9.0 fb$^{-1}$ 
of integrated luminosity taken with the CLEO II and II.V
detectors \cite{CLEOdetector} using the CESR $e^+e^-$ storage ring 
on the peak of the 
$\Upsilon(4S)$ resonance and 4.4 fb$^{-1}$ in the
continuum at 60 MeV less center-of-mass energy. The sample 
contains 19.4 million $B$ mesons.

%\section{Selection Criteria}\label{sec:Common}

Hadronic events are selected by requiring a minimum of five charged tracks,
total visible energy greater than 15\% of the center-of-mass energy, and a charged
track vertex consistent with the nominal interaction point.
To reject non-$B\overline{B}$ continuum we require that the Fox-Wolfram moment $R_2$ 
be less than 0.3 \cite{Fox-Wolf}. 
Track candidates are required to pass through a common spatial 
point defined by the luminous region. Tracks with momenta below 
900 MeV/$c$ are required to have an ionization loss in the drift 
chamber within 3 standard deviations of that expected for their 
mass hypothesis ($\pi$/K).
 Photon candidates are required
to be in the ``good barrel region," within 45$^{\circ}$ of 
the plane perpendicular to the beam line that passes through the interaction
point, and have an energy distribution in 
the CsI calorimeter consistent with that of an electromagnetic 
shower. To select $\pi^0$'s,
we require that the diphoton invariant mass be between  $-3.0$ to 
+2.5$\sigma$ of the $\pi^0$ mass, where the resolution $\sigma$ varies with momentum and has an
average value of approximately 5.5 MeV. The $\pi^0$ candidates
are then kinematically fit by constraining their invariant mass be equal to the
nominal $\pi^0$ mass.

We select $D^0$  candidates in the $K^-\pi^+$ decay mode. 
We require that the invariant mass of the $D^0$  candidates lie 
within $\pm 2.5\sigma$ of the known $D^0$ mass.  The $D^0$ mass resolution
varies with $D^0$ momentum, $p$,  as 
$p\times$0.93$\times 10^{-3}$+6.0 (units of MeV).  
We use the analogous requirement on the $D^{*0}$-$D^0$ mass
difference. In this case  
the mass difference resolution is 
0.90 MeV.

% \section{Observation of $\overline{B}^0\to D^{*0}\pi^+\pi^+\pi^-
%\pi^-$ Decays}\label{sec:Dstarp4pi}

We start by looking for the $D^{*0}(4\pi)^0$ final 
state.\footnote{In this Letter $(4\pi)^0$ will always denote the 
specific combination $\pi^+\pi^+\pi^-\pi^-$.} 
The $D^{*0}$  candidates are pooled with all combinations of
$\pi^+\pi^+\pi^-\pi^-$ mesons. 
Next, we
calculate the difference between the beam energy, $E_{beam}$, and 
the measured energy
of the five particles, $\Delta E$. The ``beam constrained" 
invariant mass of the $B$
candidates, $M_B$, is computed from the formula
$M_B^2=E_{beam}^2-(\sum_i{\bf p_i})^2.$
To further reduce backgrounds we define
\begin{equation}\label{eq:chisq}
\chi_B^2=\left({{\Delta M_{D^*}}\over {\sigma(\Delta 
M_{D^*})}}\right)^2 + 
%\left({{\Delta M_{D}}\over {\sigma(\Delta M_{D})}}\right)^2 + 
%\left({{\Delta M_{\pi^0}}\over {\sigma(\Delta 
%M_{\pi^0})}}\right)^2~~~,
%\chi_B^2=\left({{M(\pi D)-M_{D^*}}\over 
%{\sigma(M(\pi D)-M_{D^*})}}\right)^2 + 
\left({{M(K\pi)- M_{D}}\over {\sigma(M(K\pi)- M_{D})}}\right)^2 + 
\left({{M(\gamma\gamma)- M_{\pi^0}}\over 
{\sigma(M(\gamma\gamma)- M_{\pi^0})}}\right)^2~~~,
\end{equation}
where $\Delta M_{D^*}$ is the computed $D^{*0}-D^0$ mass difference minus the
nominal value and the $\sigma$'s are the measurement errors. We select candidate 
events requiring that $\chi^2_B < 5$.

%\section{Branching Fraction and $(4\pi)^0$ Mass Spectrum}

\begin{figure}[hbt]
%\vspace{-2.5cm}
%\centerline{\epsfig{figure=mb_4pi_0.eps,height=4in}\hspace{-0.9in}}
\centerline{\epsfig{figure=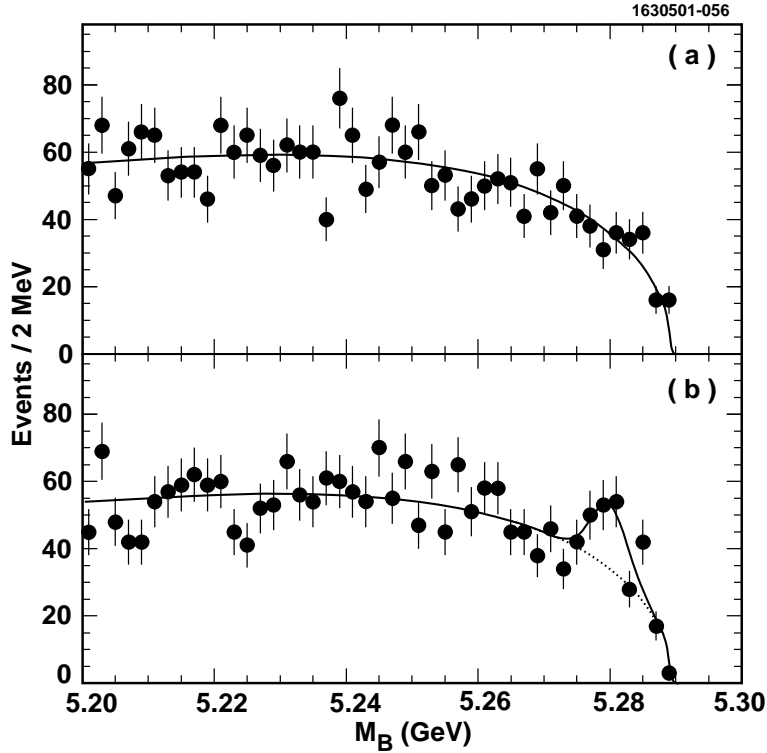,height=4in}\hspace{-0.9in}}
%\vspace{-1.0cm}
\caption{\label{mb_4pi_0}The $B$ candidate mass spectra for
the final state $D^{*0}\pi^+\pi^+\pi^-\pi^-$, (a)  
$\Delta E$ sidebands, (b) for $\Delta E$ consistent with zero.}
\end{figure}

We show the candidate $B$ mass distribution, $M_B$, for 
$\Delta E$ in the side-bands from $-6.0$ 
to $-4.0 \sigma$ and 4.0 to 6.0$\sigma$
on Fig.~\ref{mb_4pi_0}(a). The $\Delta E$ resolution
is 14 MeV ($\sigma).$ The sidebands give a good representation of the 
background in the signal region.
We fit this distribution with a shape given as 
$back(r)=p_1 r\sqrt{1-r^2}e^{-p_2(1-r^2)},$
where $r=M_B/5.2895$, and the $p_i$ are parameters given by the 
fit.

 We next view the $M_B$ distribution for events having  $\Delta E$ 
within 2$\sigma$ of zero in
Fig.~\ref{mb_4pi_0}(b). This distribution is fit with a Gaussian 
signal function of width 2.7 MeV and the background function found above 
whose 
normalization is allowed to vary. The Gaussian signal width is found from
Monte Carlo simulation. The largest and 
dominant component results from the energy spread of the beam. 
We find a total of 64$\pm16$ events, thus establishing a signal.

The error due to the background shape is evaluated in three ways.
First of all, we change the background shape by varying the
fitted parameters by 1$\sigma$. This results in a change of 
$\pm$9.3\%. Secondly,
we allow the shape, $p_2$, to vary (the normalization, $p_1$, was 
already
allowed to vary). This results in 11\% decrease in the number of 
events.
Finally, we choose a different background function,
$back'(r)=p_1 r\sqrt{1-r^2}\left(1+p_2 r + p_3 r^2 +p_4 r^3 
\right),$
and repeat the fitting procedure. This results in a 9.3\% 
increase in the
number of events. We assign $\pm$7 event systematic error due to this source. 

We have investigated two mode-specific backgrounds that could, in principle, 
induce fake signals. These include the final states 
$D^{*+}\pi^+\pi^-\pi^-\pi^0$ where we miss the slow $\pi^+$ from the $D^{*+}$ 
decay and the $\pi^0$ and $D^0$ happen to satisfy the $D^{*0}$ requirement, and
$D^0\pi^+\pi^+\pi^-\pi^-\pi^0$, where the $D^0$ and the $\pi^0$ happen to 
satisfy the $D^{*0}$ requirement. We find that the efficiencies for each of 
these modes to contribute are small. 
The first final state was measured as having a branching ratio of 1.72\% 
\cite{Wangsbest}. It would contribute $0.4 \pm 0.5$ events. The second final 
state has never been measured. It would contribute $1.6 \pm 0.5$ events per 1\% 
branching ratio. Taking into account all sources of systematic error we 
observe 64$\pm$16$\pm$7 signal events.

\begin{figure}[htb]
%\vspace{-2.5cm}
%\centerline{\epsfig{figure=beff_4pi.eps,height=3.5in}}
\centerline{\epsfig{figure=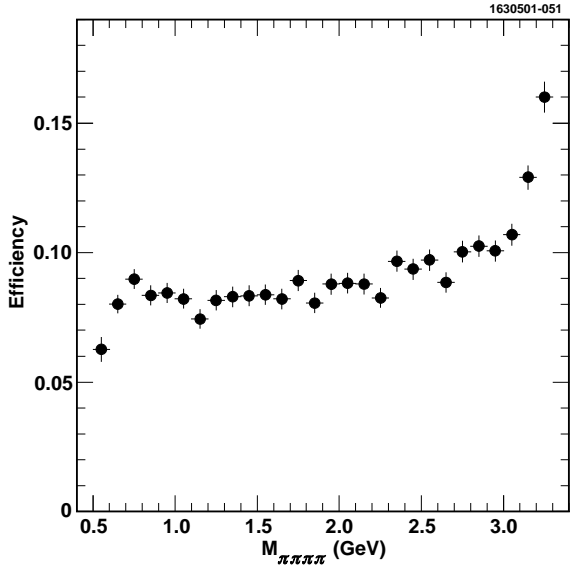,height=3.0in}}
%\vspace{0.3cm}
\caption{ \label{beff_m4pi}The efficiency for the final state
$D^{*0}\pi^+\pi^+\pi^-\pi^-$.}
\end{figure}

\begin{figure}[bt]
%\vspace{-2.5cm}
%\centerline{\epsfig{figure=m4pi.eps,height=4in}}
\centerline{\epsfig{figure=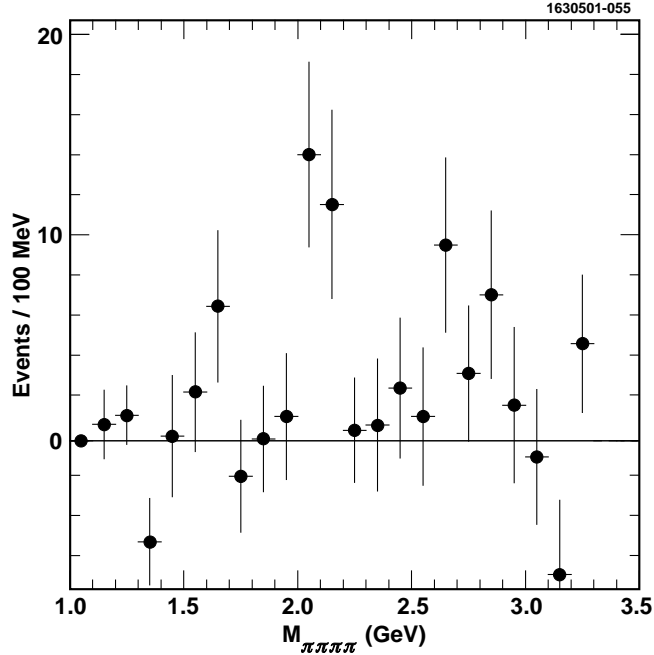,height=3.5in}}
%\vspace{-0.2cm}
\caption{ \label{m4pi}The invariant mass spectrum of 
$\pi^+\pi^+\pi^-\pi^-$ for the final state
$D^{*0}\pi^+\pi^+\pi^-\pi^-$, found by 
fitting the $B$ yield in bins of 4$\pi$ mass.}
\end{figure}

In order to find the branching ratio we use the Monte Carlo-determined
efficiency, shown in Fig.~\ref{beff_m4pi} as a 
function of $(4\pi)^0$ mass. 
Since the efficiency varies with mass we need to determine the 
$(4\pi)^0$ mass spectrum in order to determine the branching ratio.
To rid ourselves of the problem of the background shape, we fit 
the $B$ candidate mass spectrum in 100 MeV bins of $(4\pi)^0$ mass. 
The resulting $(4\pi)^0$ mass 
spectrum is shown in Fig.~\ref{m4pi}.
We find
\begin{equation}
{\cal B}(\overline{B}^0\to D^{*0}\pi^+\pi^+\pi^-\pi^-)
=(0.30\pm 0.07\pm 0.06)\%~~~.
\end{equation}
The systematic error arises mainly from our lack of knowledge 
about the 
tracking and $\pi^0$ efficiencies. We assign errors of $\pm$2.2\% 
on the efficiency of each charged track, and $\pm$5.4\% for the $\pi^0$.
The
total tracking error is found by adding the error in the charged particle
track finding efficiency linearly for the 6 ``fast" charged tracks and then
in quadrature with the slow pion from the $D^{*0}$ decay.
The error on the background shape has been
discussed above. We take a conservative estimate of the systematic 
error due to this source of $\pm$11\%. We use the current 
Particle Data Group
values for the relevant $D^{*0}$ and $D^0$ branching ratios of
(61.9$\pm$2.9)\% ($D^{*0}\to\pi^0 D^0$) and 
(3.83$\pm$0.09)\% ($D^0\to K^-\pi^+$) \cite{PDG}. 
The relative
errors, 4.7\% for the $D^{*0}$ branching ratio and 2.3\% for the 
$D^0$ are
added in quadrature to the background shape error, the $\pi^0$ detection 
efficiency uncertainty and the tracking error. The
total positive systematic error is 19\%. We also allow for cross-feed backgrounds amounting to 4 events giving a total negative systematic
error of 20\%.

\begin{figure}[htb]
%\vspace{-2.5cm}
%\centerline{\epsfig{figure=dspi.eps,height=4in}}
\centerline{\epsfig{figure=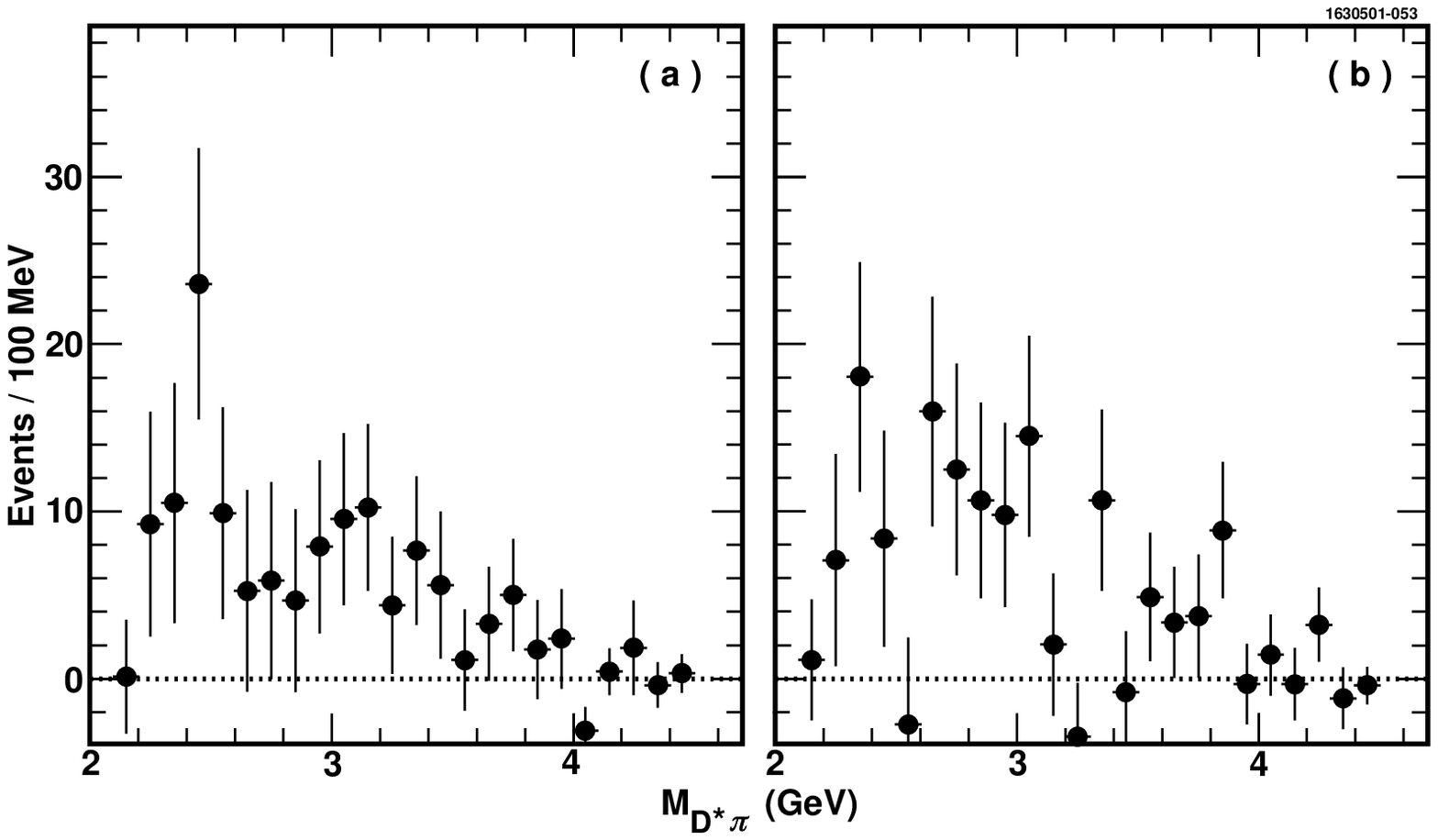,height=3.1in}}
\vspace{0.2cm}
\caption{ \label{dspi}The invariant mass spectra of 
(a) $D^{*0}\pi^+$, and (b) $D^{*0}\pi^-$ for the final state 
$D^{*0}\pi^+\pi^+\pi^-\pi^-$, found by 
fitting the $B$ yield in bins of $D^{*0}\pi^{\pm}$ mass. (There are two
combinations per event.)}
\end{figure}

%\section{Properties of the Final State}
In Fig.~\ref{dspi}(a) we show the $D^{*0}\pi^+$ invariant mass spectrum, obtained
by fitting the number of $\overline{B}^0$ events as a function of $D^{*0}\pi^+$
mass. (There are two combinations per event.) We also show for comparison
the $D^{*0}\pi^-$ mass spectrum, where no structure is expected. We see evidence for an
excess of events in the region of the $D^{**+}(2420)$ and $D^{**+}(2460)$.
There are four $D^{**}$ mesons. Two have relatively narrow widths and decay
into $D^*\pi$, whereas
two are wide, with only one decaying into $D^*\pi$ \cite{Ddoublestar}. It is difficult to quantitatively evaluate the fraction
of total $D^{**+}$ production in our data. We find that $\sim$70\% of the signal
has one mass combination between 2.3 and 2.6 GeV.
% In fact all of this final state is consistent with arising from
% $D^{**+}\pi^+\pi^-\pi^-$ production. 

\begin{figure}[htb]
%\vspace{-2.5cm}
%\centerline{\epsfig{figure=m3pi_d2select.eps,height=4in}}
\centerline{\epsfig{figure=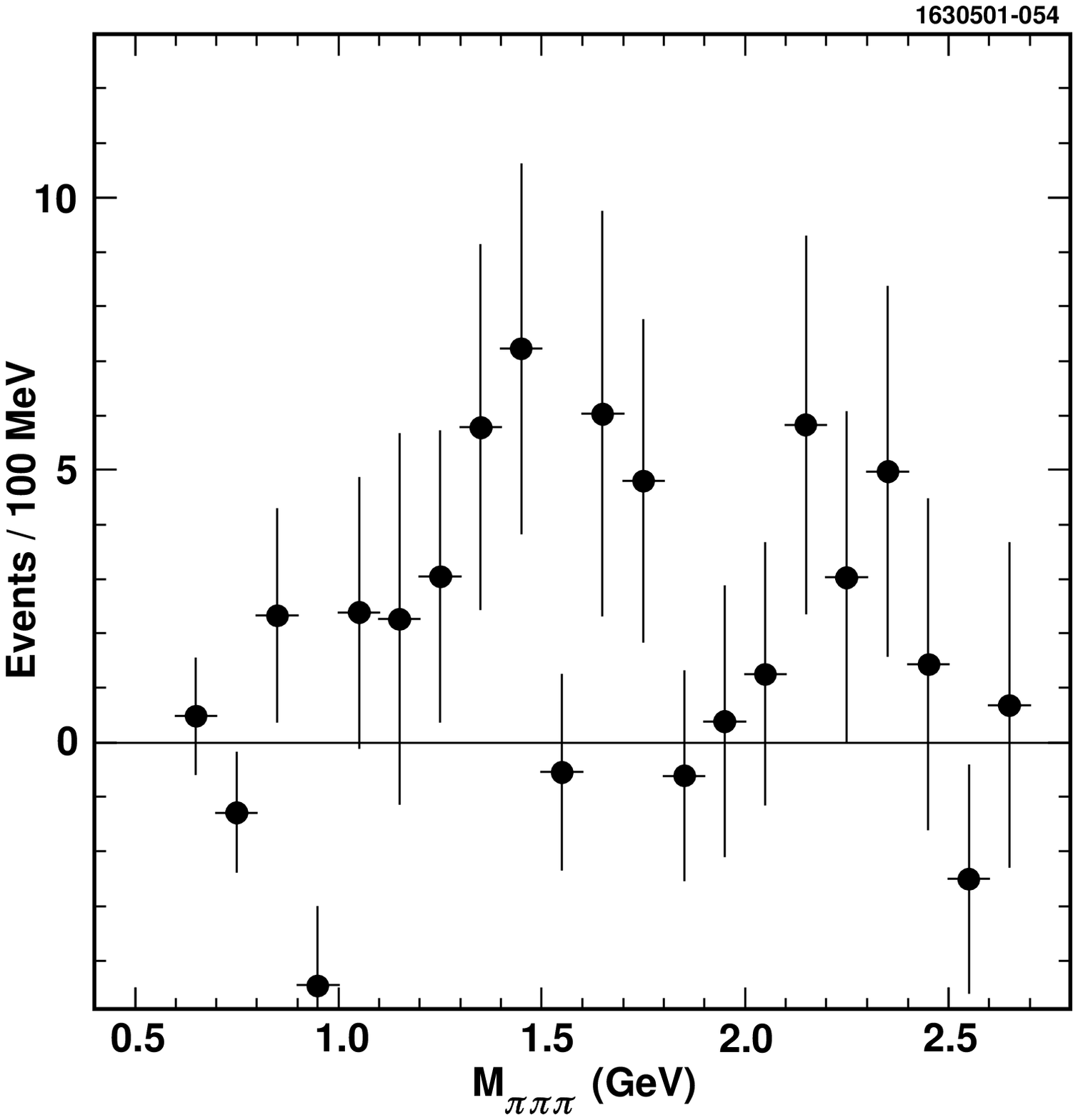,height=3in}}
\vspace{0.2cm}
\caption{ \label{m3pi_d2select}The invariant mass spectra of 
$\pi^+\pi^-\pi^-$ when the $D^{*0}\pi^+$ invariant mass is
between 2.3 and 2.6 GeV, found by fitting the $B$ yield in bins
of $\pi^+\pi^-\pi^-$ mass.}
\end{figure}

In Fig.~\ref{m3pi_d2select} we show the $\pi^+\pi^-\pi^-$ mass spectrum when
the $D^{*0}\pi^+$ mass is required to be between 2.3 and 2.6 GeV. Here 
we fit the $B$ yield as a function of $\pi^+\pi^-\pi^-$ mass.
There is no clear feature present.

\begin{figure}[htb]
%\vspace{-2.5cm}
%\centerline{\epsfig{figure=dg_dm2_1.eps,height=4in}}
\centerline{\epsfig{figure=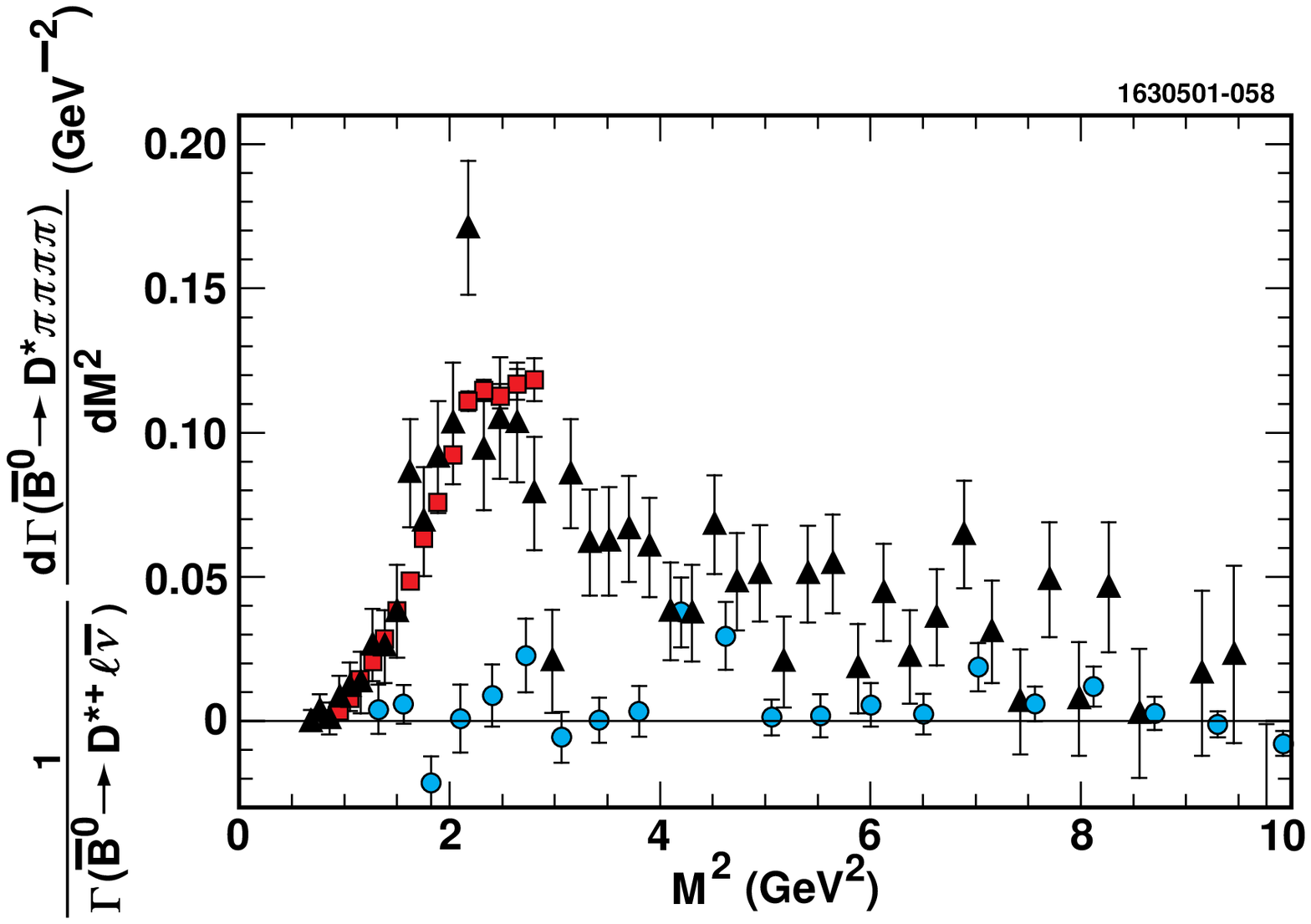,height=4in}}
\vspace{0.2cm}
\caption{ \label{dg_dm2_1}${d\Gamma\over dM^2}\left(\overline{B^0}\to 
D^*\pi\pi\pi\pi\right)$, where $M^2$ is the 4$\pi$ invariant mass squared,
normalized to the semileptonic width $\Gamma(\overline{B^0}\to D^*\ell^-
\bar{\nu})$. The circles are the CLEO data for $\overline{B}^0\to 
D^{*0}\pi^+\pi^+\pi^-\pi^-$, the squares the model prediction of LLW and
the triangles are the data for $\overline{B}^0\to D^{*+}\pi^+\pi^-\pi^-\pi^0$.
There is an additional systematic normalization uncertainty on the triangle 
points of 9\% and on the circles of 16\%.}
\end{figure}

Let us now see how the presence of this final state affects the LLW prediction.
In Fig.~\ref{dg_dm2_1} we show the CLEO data \cite{Wangsbest}
for ${d\Gamma\over dM^2}\left(\overline{B}^0\to 
D^{*+}\pi^+\pi^-\pi^-\pi^0\right)$ plotted as a function of
the four-pion invariant
mass squared, normalized to the semileptonic rate \cite{PDG}, and compared with the 
LLW prediction \cite{LLW}. We also plot 
${d\Gamma\over dM^2}\left(\overline{B}^0\to D^{*0}\pi^+\pi^+\pi^-\pi^-\right)$,
again normalized to the semileptonic rate.
In principle a non-zero rate in the $D^{*0}(4\pi)^0$ final state
is indicative of an additional
contribution to the $D^{*+}(4\pi)^-$ final state, beyond what is expected
in factorization, and needs to be subtracted
to make an accurate prediction.\footnote{If the $D^{*0}(4\pi)^0$
final state comes from $D^{**+}$ production, then some $D^{**+}$,
(half as much according to isotopic-spin symmetry) should be present
in the $D^{*+}(4\pi)^-$.}
 In fact, the $D^{*0}(4\pi)^0$ rate is 
consistent with zero in the
mass squared region covered by the LLW prediction.

%\section{Conclusions}

In conclusion we have made the first measurement of 
${\cal B}\left(\overline{B}^0\to D^{*0}\pi^+\pi^+\pi^-\pi^-\right)
=(0.30 \pm 0.07 \pm 0.06)$\%. 
The reaction has a large component of $D^{**+}\to
D^{*0}\pi^+$. We determine the relative rate 
\begin{equation}
R_{0-}={{{\Gamma}\left(\overline{B}^0\to D^{*0}\pi^+\pi^+\pi^-\pi^-\right)}
\over {
{\Gamma}\left(\overline{B}^0\to D^{*+}\pi^+\pi^-\pi^-\pi^0\right)}}
= 0.17\pm0.04\pm0.02
~~.
\end{equation}
We have no evidence in the $D^{*0}$ final state for $4\pi^0$ masses below
2.9 GeV$^2$ and set an upper limit on $R_{0-}<0.13$ at 90\% confidence level in this
restricted mass region.

LLW have used the $\overline{B}^0\to D^{*+}\pi^+\pi^-\pi^-\pi^0$ reaction
to test the 4$\pi$ mass dependence of factorization. They point out that
a perturbative origin for factorization should cause a weakening of the
prediction with increasing 4$\pi$ mass. However if the basis for factorization
is the large $N_c$ limit, where $N_c$ refers to the
number of colors, no such weakening should occur.  
LLW also suggest that the presence of $D^{**}$ production might cause their 
factorization test 
to be inaccurate. We have found such a presence in the analogous reaction
$\overline{B}^0\to D^{*0}\pi^+\pi^+\pi^-\pi^-$. However, the 4$\pi$ mass
region that is populated is higher than that used by LLW, so no effect on
their prediction can be inferred.

%\section{Acknowledgements}
We thank Z. Ligeti, M. Luke and M. Wise for interesting discussions.
We gratefully acknowledge the effort of the CESR staff in providing us with
excellent luminosity and running conditions.
This work was supported by 
the National Science Foundation,
the U.S. Department of Energy,
the Research Corporation,
the Natural Sciences and Engineering Research Council of Canada
and the Texas Advanced Research Program.

\end{document}